\documentclass{cpcauth}

\usepackage{amssymb}

\begin{document}

\begin{frontmatter}



\title{Multicanonical Simulations Step by Step} 


\author{Bernd A. Berg {\small (berg@hep.fsu.edu)} }

\address{ Department of Physics, Florida State University, 
Tallahassee FL 32306, USA}

\date{August 9 2002}

\begin{abstract}
The purpose of this article is to provide a starter kit for multicanonical 
simulations in statistical physics.  Fortran code for the $q$-state Potts 
model in $d=2,\,3,\dots$ dimensions can be downloaded from the Web and this 
paper describes simulation results, which are in all details reproducible 
by running prepared programs. To allow for comparison with exact results,
the internal energy, the specific heat, the free energy and the entropy 
are calculated for the $d=2$ Ising ($q=2$) and the $q=10$ Potts model.
Analysis programs, relying on an all-log jackknife technique, which is 
suitable for handling sums of very large numbers, are introduced to 
calculate our final estimators.
\end{abstract}

\end{frontmatter}

\section{Introduction} \label{sec_intro}

The conventional Metropolis~\cite{Me53} method simulates the Gibbs
canonical ensemble at a fixed temperature $T$ and allows for easy
calculations of the (internal) energy and functions thereof. However,
some of the most important quantities of statistical physics, free
energy and entropy, can only be obtained by tedious integrations.
One way to overcome this problem is by multicanonical (MUCA) 
simulations, which calculate canonical expectation values over a 
temperature range in a single simulation by using the weight
factor~\cite{BeNe92}
\begin{equation} \label{wspectral}
w_{1/n} = {1\over n(E)} = e^{-S(E)} =
e^{-b(E)E-a(E)} 
\end{equation}
where $n(E)$ is the number of states with energy $E$ and $S(E)$ the
microcanonical entropy. In an extension of the microcanonical
terminology on may call $b(E)$ microcanonical inverse temperature 
($b(E)= 1/T(E)$ in natural units with Boltzmann constant $k_B=1$) 
and $a(E)$ microcanonical, dimensionless free energy, see
appendix~\ref{Recursion}. 

MUCA simulations became popular with the interface tension 
calculation~\cite{BeNe92} of the $2d$ 10-state Potts model, when the 
method emerged as the winner of largely disagreeing estimates, which 
after their publication became resolved by exact values~\cite{BoJa92}. 
Similar simulation concepts can actually be traced back to the work by 
Torrie and Valleau~\cite{ToVa77} in the 1970s. In recent years the MUCA 
method has found many applications, besides for first order phase 
transitions mainly for complex systems including spin glasses and 
peptides, see~\cite{Be01} for a brief review and a summary of related 
methods.

The scope of this article is limited to the Ising model and its 
generalization in form of $q$-state Potts models, for a review 
see~\cite{Wu82}.  Fortran routines which work in arbitrary integer
dimensions $d=2,\,3,\dots$ are provided, but we confine our 
demonstrations to $d=2$, to allow for 
comparison with rigorous analytical calculations. The Ising 
model simulation is seen to match the exact finite lattice results of 
Ferdinand and Fisher~\cite{FeFi69}, while for the $q=10$ Potts model 
one finds agreement with the rigorously known transition temperature 
and latent heat of Baxter~\cite{Ba73}. Details of the model are summarized 
in the first part of section~\ref{Preliminaries}. In the second part
of this section the downloading of the Fortran code and its use are
explained.  

In section~\ref{MUCA} MUCA simulations are treated. The temperature 
dependence of the standard thermodynamic quantities -- energy, specific 
heat, free energy and entropy -- is calculated for the $2d$ Ising 
model as well as for the $2d$ 10-state Potts model and the canonically
re-weighted histograms are shown. Special attention is given to the
analysis procedure, which has to be able to handle sums of very
large numbers. This is done by using only the logarithms until finally 
the quotient of two such numbers is obtained. 
Jackknife~\cite{Mi74} binning is used to minimize bias problems
which occur in the re-weighting of the simulation data to canonical
ensembles. 

Using the provided Fortran code and following the 
instructions allows for a step by step reproduction of the figures 
and all other numerical results presented in this article. This could 
be a desirable standard for more involved simulations too.
Some conclusions are given in the final section~\ref{conclusions}.

\section{Getting Started} \label{Preliminaries}

\subsection{The Potts model} \label{Potts}

We introduce the Potts models on $d$-dimensional hypercubic lattices 
with periodic boundary conditions. For this paper we stay close to the 
notation used in the accompanying computer programs and define the
energy $E$ via the action variable
\begin{equation} \label{iact}
 {\tt iact} = \sum_{<ij>} \delta (q_i^{(k)},q_j^{(k)})\,,\ \
 E = {2 d\,N\over q} - 2\,{\tt iact}
\end{equation}
where $\delta (q_i,q_j)$ is the Kronecker delta. The sum $<ij>$ is 
over the nearest neighbor lattice sites and $q_i^{(k)}$ is the 
{\it Potts state} of configuration $k$ at site $i$. For the 
$q$-state Potts model $q^{(k)}_i$ takes on the values $1,\dots ,q$.  
As the variable {\tt iact} takes on integer values, it allows for 
convenient histograming of its values during the updating process.
Occasionally, we use the related mean values
\begin{equation} \label{actm}
{\tt actm} = {\tt iact}\, /\, (d\, N) ~~{\rm and}~~
e_s = E\, /\, N\ .
\end{equation}
Each configuration (microstate of the system) $k$ defines a
particular arrangements of all states at the sites and, vice versa,
each arrangement of the states at the sites determines uniquely a
configuration:
\begin{equation} \label{k}
 k = \{ q_1^{(k)}, \dots , q_N^{(k)} \} \ . 
\end{equation}
The expectation value of an observable $O$ is defined by
\begin{equation} \label{O}
\langle O\rangle = Z^{-1} \sum_{k=1}^K O^{(k)}\, e^{-\beta\,E^{(k)}}
\end{equation}
where the sum is over all microstates and the partition function 
$Z=Z(\beta )$ normalizes the expectation value of the unit operator 
to $\langle 1\rangle = 1$. As there are $q$ possible Potts states at 
each site, the total number of microstates is
\begin{equation} \label{K}
 Z(\beta =0) = q^N\ .  
\end{equation}
Including $\beta=0$ in a MUCA simulation allows for the normalization
of the partition function necessary to calculate the canonical free 
energy and the entropy as a function of the temperature.

Our definition(\ref{O}) of $\beta$ agrees with the one commonly used for 
the Ising model~\cite{Hu87}, but disagrees by a factor of two with the 
one used for the Potts model in~\cite{Ba73,BoJa92}:
\begin{equation} \label{beta_def}
\beta\ =\ \beta^{\rm Ising}\ = \beta^{\rm Potts}\, /\, 2\, .
\end{equation}
For the $2d$ Potts models a number of exact results are known in the
infinite volume limit. The critical temperature~\cite{Ba73} is
\begin{equation} \label{2d_Potts_bc}
{1\over 2}\, \beta_c^{\rm Potts} = \beta_c = {1\over T_c} 
= {1\over 2}\, \ln (1 + \sqrt{q} ), ~~q=2,3,\dots\ .
\end{equation}
The phase transition is second order for $q\le 4$ and first order for 
$q\ge 5$. At $\beta_c$ the average energy per Potts state is~\cite{Ba73}
\begin{equation} \label{Potts_es}
 - e_{s}^c\ =\ 2 + 2/\sqrt{q} 
\end{equation}
where, by reasons of consistency with the Ising model notation, also
our definition~(\ref{actm}) of $e_{s}$ differs by factor of two from 
the one used in most Potts model literature. For the 
first order transitions at $q\ge 5$ equation (\ref{Potts_es}) gives the
average of the limiting energies from the ordered and the disordered
phase. The exact infinite volume latent heat $\triangle e_s$
and the entropy jumps $\triangle s$ were also calculated by
Baxter~\cite{Ba73}, whereas the  interfacial tensions $f_s$ were
derived more recently~\cite{BoJa92}.

\subsection{The Fortran Code} \label{Fortran}

\begin{figure}[t]
 \begin{picture}(100,100)
    \put(0, 0){\includegraphics{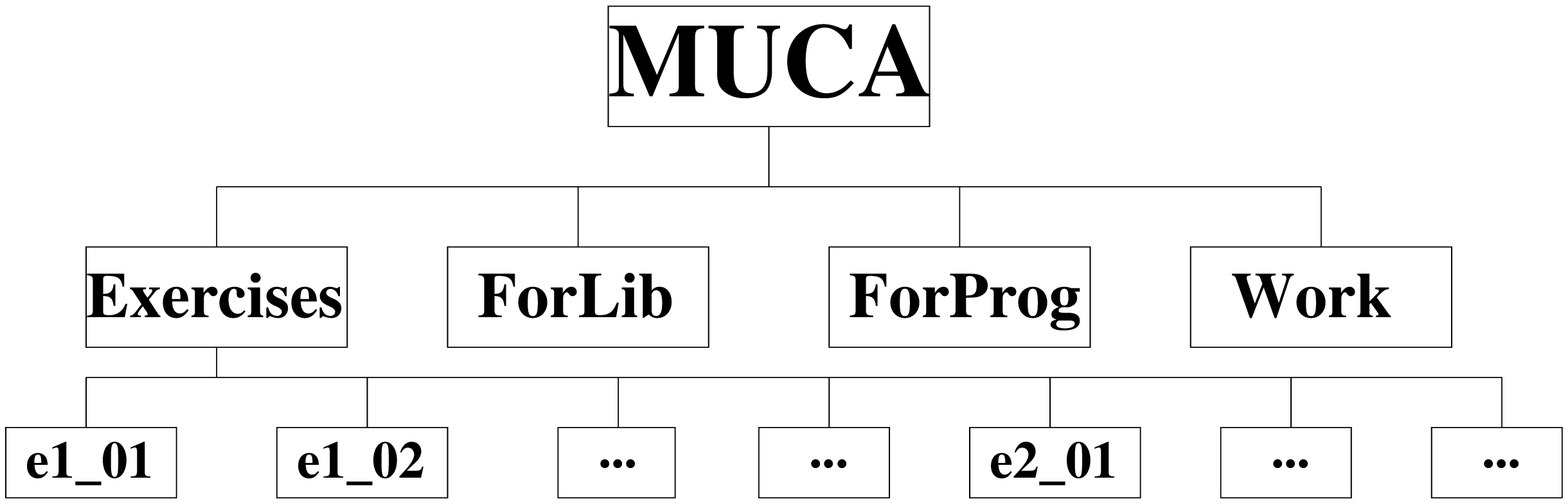}}
  \end{picture}
\caption{Fortran code directory structure. \label{fig_fort} }
\end{figure}

Figure~\ref{fig_fort} shows the directory tree in which the Fortran 
routines are stored. {\tt MUCA} is the parent directory and on the 
first level we have the directories {\tt Exercises},
{\tt ForLib}, {\tt ForProg} and {\tt Work}. The master
code is provided in the directories {\tt ForLib} and {\tt ForProg}.
{\tt ForLib} contains the source code of a library of functions
and subroutines. The library is closed in the sense that no reference
to non-standard functions or subroutines outside the library is ever
made.  The master versions of the main programs and certain routines 
(which need input from the parameter files discussed below) are
contained in the subdirectory {\tt ForProg}. The demonstrations of 
this article are contained in the subdirectories of {\tt Exercises}.

To download the code, start with the URL 
{\tt www.hep.fsu.edu/\~\,}$\!${\tt berg} and click the {\tt Research}
link, then the link {\tt Multicanonical Simulations}. On this page
follow the link {\tt Fortran Code} and get either the file
{\tt muca.tar} ($399\,$KB) or the file {\tt muca.tgz} ($40\,$KB).
On most Unix platforms you obtain the directory structure of 
figure~\ref{fig_fort} from {\tt muca.tar} by typing
\begin{equation} \label{tar}
 {\tt tar}\ {\tt -\! xvf}\ {\tt muca.tar} 
\end{equation}
alternatively from {\tt muca.tgz} by typing either 
{\tt tar -zxvf muca.tgz} or {\tt gunzip muca.tgz} followed 
by (\ref{tar}).

You obtain the results of this paper by compiling and running the 
code prepared in subdirectories of {\tt Exercises}, e.g. 
{\tt f77 -O program.f} followed by {\tt ./a.out}. Due to the 
{\tt include} and {\tt parameter} file structure used, the programs 
and associated routines of {\tt ForProg} compile only in 
subdirectories which are two levels down from the {\tt MUCA} parent 
directory.  This organization should be kept, unless you have strong 
reasons to change the dependencies. The present structure allows to 
create {\tt Work} directories for various projects, with the actual 
runs done in the {\tt Work} subdirectories. Note that under MS Windows 
with the (no longer marketed) MS Fortran compiler a modification of the 
{\tt include} structure of the code turned out to be necessary.
If such problems are encountered, one solution is to copy all needed 
files to the subdirectory in question and to modify all 
include statements accordingly.

Each {\tt Exercises} subdirectory contains a 
\begin{equation} \label{readme}
 {\tt readme.txt} 
\end{equation}
file with instructions, which should be followed. The subdirectories
{\tt e1}$\dots$ prepare some code to check for the correctness of
the conventional Metropolis code and the subdirectories {\tt e2}$\dots$
prepare examples of multicanonical simulations, which are discussed
in the next section.
The simulation parameters are set in the files
\begin{equation} \label{parameters}
{\tt lat.par}\,,\ {\tt lat.dat}\,,\ {\tt potts.par}\,,\
{\tt mc.par}~~{\rm and}~~{\tt muca.par}\,,
\end{equation}
or a subset thereof, which are also kept in each of the subdirectories 
of {\tt Exercises}. The dimension {\tt nd} of the system and the maximum 
lattice length {\tt ml} are defined in {\tt lat.par}. In {\tt lat.dat} 
the lattice lengths for all directions are assigned to the array 
{\tt nla}, which is of dimension {\tt nd}. This allows for asymmetric 
lattices.  The number of Potts states, {\tt nq}, is defined in 
{\tt potts.par}.  The parameters of the conventional Metropolis 
simulation are defined in {\tt mc.par}: These are the $\beta$ value 
{\tt beta}, which defines the initial weights in case of a MUCA 
simulation, the number of equilibrium 
sweeps {\tt nequi}, the number of measurement repetitions {\tt nrpt} 
and the number of measurement sweeps {\tt nmeas}. Additional parameters 
of the MUCA recursion are defined in {\tt muca.par}: The maximum number 
of recursions {\tt nrec\_max}, the number of sweeps between
recursions {\tt nmucasw} and the maximum number of 
tunnelings~(\ref{muca_tunneling}) {\tt maxtun}, which terminates the 
recursion unless {\tt nrec\_max} is reached first.

Whenever data for a graphical presentation are generated by our code,
it is in a form suitable for {\it gnuplot}, which is freely available.
Gnuplot driver files {\tt fln.plt} are provides in the solution 
directories, such that one obtains the the plot by typing 
{\tt gnuplot\ fln.plt} on Unix and Linux platforms (under MS Windows 
follow the gnuplot menu).

\section{Multicanonical Simulations} \label{MUCA}
 
A conventional, canonical simulation 
calculates expectation values at a fixed temperature $T$ and can, by 
re-weighting techniques, only be extrapolated to a vicinity of this 
temperature~\cite{FeSw88}.  In contrast, a single MUCA 
simulation allows to obtain equilibrium properties of the Gibbs ensemble
over a range of temperatures, which would require many canonical 
simulations. This coined the name multicanonical. The MUCA method 
requires two steps:

\begin{enumerate}

\item Obtain a \textit{working estimate} $w_{mu}(k)$ of the weights 
      $w_{1/n}(k)$. Working estimate means that the approximation 
      to~(\ref{wspectral}) has to be good enough to ensure movement
      in the desired energy range, but deviations of $w_{mu} (E)$
      from (\ref{wspectral}) by a factor of, say, ten are tolerable.

\item Perform a Markov chain MC simulation with the final, fixed 
      weights $w_{mu}(k)$. Canonical expectation values 
      are found by re-weighting to the Gibbs ensemble.

\end{enumerate}

To obtain working estimates $w_{mu}(k)$ of the weight factors 
(\ref{wspectral}), a slightly modified version of the recursion of 
Ref.~\cite{Be96} is used. As the analytical derivation of the 
modified recursion has so far only been published in conference 
proceedings, it is for the sake of completeness included in 
appendix~\ref{Recursion} of this paper. The Fortran implementation
is given by the subroutine 
\begin{equation} \label{p_mu_rec}
{\tt p\_ mu\_ rec.f}
\end{equation}
of {\tt ForProg}. One subtlety is that two histogram arrays {\tt hup} 
and {\tt hdn} are introduced to keep separately track of the use of 
upper and lower entries of nearest neighbor pairs. Detailed 
explanations of the code will be part of a book~\cite{Berg}.

The question whether more efficient recursions exists is far from
being settled. For instance, F. Wang and Landau~\cite{WaLa01} made
recently an interesting proposal. Exploratory comparisons with the
recursion used in the present paper reveal similar 
efficiencies~\cite{Ok02}.

In between the recursion steps the Metropolis updating routine
\begin{equation} \label{potts_met}
{\tt potts\_ met\_ f}
\end{equation}
is called, 
which implements the standard Metropolis algorithm~\cite{Me53} for 
general weights. The random number generator of Marsaglia et 
al.\cite{MaZa90} is integrated to ensure identical results on 
distinct platforms.

\subsubsection{Example runs}

First, we illustrate the MUCA recursion for the $20^2$ Ising 
model. We run the recursion in the range
\begin{equation} \label{2dI_namin_max}
{\tt namin} = 400 \le {\tt iact} \le 800 = {\tt namax}\ . 
\end{equation}
These values of
{\tt namin} and {\tt namax} are chosen to cover the entire range
of temperatures, from the completely disordered ($\beta =0$) region
to the groundstate ($\beta\to\infty$). In many applications the actual 
range of physical interest is smaller, {\tt namin} and {\tt namax} 
should correspondingly be adjusted, because the recursion time
increases quickly with the range. The recursion is completed after
{\tt maxtun} tunneling events have been performed. A {\it tunneling 
event} is defined as an updating process which finds its way from
\begin{equation} \label{muca_tunneling}
{\tt iact} = {\tt namin} ~~~{\rm to}~~~ {\tt iact} = {\tt namax}
~~~{\rm and}~~{\rm back}\ . 
\end{equation}
This notation comes from the applications
of the method to first order phase transitions~\cite{BeNe92}, for which
{\tt namin} and {\tt namax} are separated by a free energy barrier 
in the canonical ensemble.  Although the MUCA method removes
this barrier, the terminus tunneling was kept.  The requirement that 
the process tunnels also back is included in the definition, because a 
one way tunneling is not indicative for the convergence of the recursion.
Most important, the process has still to tunnel when the weights
are frozen for the second stage of the simulations Note that things
work differently for the Wang-Landau recursion~\cite{WaLa01}. It has 
no problems to tunnel in its initial stage, but its estimates of the 
spectral density are still bad, such that the tunneling process gets
stuck as soon as the weights are fixed.

For most applications ten tunnelings during our recursion part
lead to acceptable weights. If the requested number of tunnelings
is not reached after a certain maximum number of recursion steps, 
the problem will disappear in most cases by
rerunning (eventually several times) with different 
random numbers. Otherwise, the number of sweeps between
recursions should be enlarged, because our recursion is strictly  
only valid when the system is in equilibrium. One may even
consider to discard some sweeps after each recursion step
to reach equilibrium, but empirical evidence indicates that
the improvement (if any) does not warrant the additional CPU time. 
The disturbance of the equilibrium is weak when the weight function 
approaches its fixed point. In the default setting of our programs
we take the number of sweeps between recursion steps inversely 
proportional to the acceptance rate, because 
equal number of accepted moves is
a better relaxation criterion than an equal number of sweeps.

In the subdirectory {\tt e2\_01} of {\tt Exercises} a $20\times 20$ 
lattice Ising model simulation is prepared for which we requested 
ten tunneling events. We we find them after 787 recursions and 64,138 
sweeps, corresponding to an average acceptance rate of 
{\tt 20*787/64138=0.245} (the acceptance rate can be calculated
this way, because the number of accepted sweeps triggers the recursion). 
Almost half of the sweeps are spent to achieve the first tunneling event.
Subsequently, an MUCA production run of 10,000 equilibrium and
$32\times 10,000$ sweeps with measurements is carried out. On a GHz
Linux PC the entire runtime (recursion plus production) is about
thirty seconds. In the subdirectory {\tt e2\_02} 
a similar simulation is prepared for the $2d$ 10-state Potts model 
on a $20^2$ lattice.

\subsection{Re-weighting to the canonical ensemble}

Let us assume that we have performed a MUCA simulation
which covers the action histogram needed for a temperature range
\begin{equation} \label{beta_range}
\beta_{\min}\ \le\ \beta\le\ \beta_{\max}\ .
\end{equation}
In practice this means that the parameters {\tt namax} and {\tt namin} 
in {\tt muca.par} have to be chosen such that 
$$ {\tt namin} \ll \overline{\rm act}(\beta_{\min}) ~~{\rm and}~~
 \overline{\rm act}(\beta_{\max}) \ll {\tt namax} $$
holds, where $\overline{\rm act}(\beta)$ is the canonical expectation
value of the action variable~(\ref{iact}).
The $\ll $ conditions may be relaxed to equal signs, if 
$b({\tt iact}) = \beta_{\min}$ is used for all action values 
${\rm iact}\le\overline{\rm act}(\beta_{\min})$
and $b({\tt iact}) = \beta_{\max}$ for all action values 
${\rm iact}\ge \overline{\rm act}(\beta_{\max})$. 

Given the MUCA time series, where $i=1,\dots,n$ labels the generated 
configurations, the definition~(\ref{O}) of the canonical expectation 
values leads to the MUCA estimator
\begin{equation} \label{O_muca}
\overline{O} = 
\end{equation}
$$ { \sum_{i=1}^n O^{(i)}\, \exp \left[ -\beta\,E^{(i)} + 
b(E^{(i)})\,E^{(i)}-a(E^{(i)}) \right] \over \sum_{i=1}^n \exp 
\left[ -\beta\,E^{(i)} + b(E^{(i)})\,E^{(i)}-a(E^{(i)}) \right] } .$$
This formula replaces the MUCA weighting of the simulation by 
the Boltzmann factor of equation~(\ref{O}). The denominator differs 
from $Z$ by a constant factor, which drops out because the numerator 
differs by the same constant factor from the numerator of~(\ref{O}).
If only functions of the energy (in our computer programs the action 
variable) are calculated, it is sufficient to keep 
histograms instead of the entire time series. For an operator 
$O^{(i)} = f(E^{(i)})$ equation~(\ref{O_muca}) simplifies 
then to
\begin{equation} \label{f_muca}
\overline{f} = 
\end{equation}
$$ { \sum_E f(E)\, h_{mu}(E)\, \exp 
\left[ -\beta\,E + b(E)\,E-a(E) \right] \over \sum_E h_{mu}(E)\, 
\exp \left[ -\beta\,E + b(E)\,E-a(E) \right] } .$$
where $h_{mu}(E)$ is the histogram sampled during the MUCA
production run and the sums are over all energy values for which
$h_{mu}(E)$ has entries. When calculating error bars for estimates
from equations~(\ref{O_muca}) or~(\ref{f_muca}), we employ 
jackknife~\cite{Mi74} estimators to reduce bias problems.

A computer implementation of equations~(\ref{O_muca}) 
and~(\ref{f_muca}) requires care. The differences between the largest
and the smallest numbers encountered in the exponents can be really
large. To give one example, for the Ising model on a $100\times 100$ 
lattice and $\beta = 0.5$ the groundstate configuration contributes
$-\beta E = 10^4$, whereas for a disordered configuration $E=0$ is 
possible. Clearly, overflow disasters will result, if we ask Fortran 
to calculate numbers like 
$\exp (10^4)$.  When the large terms in the numerator and denominator 
take on similar orders of magnitude, one can avoid them by subtracting 
a sufficiently large number in all exponents of the numerator as well 
as the denominator, resulting in a common factor which divides out.
Instead of overflows one encounters harmless underflows of the type 
$\exp (-10^4)$. We implement the
idea in a more general fashion, which remains valid when 
the magnitudes of the numerator and the denominator disagree. We 
avoid altogether to calculate large numbers and deal only with the 
logarithms of sums and partial sums. 

We first consider sums of positive numbers and discuss the 
straightforward generalization to arbitrary signs afterwards. For 
$C=A+B$ with $A>0$ and $B>0$ we calculate $\ln C = \ln (A+B)$ from 
the values $\ln A$ and $\ln B$, without ever storing either $A$ or $B$ 
or $C$. The basic observation is that
\begin{eqnarray} \label{lnC}
\ln C &=& \ln \left[ \max (A,B)\,\left( 1+{\min (A,B)\over \max (A,B)}
\right) \right] \\ \nonumber
&=& \max \left( \ln A,\ln B \right) + \\ \nonumber
\ln \{ \, 1 &+& \exp \left[ \min (\ln A,\ln B)
- \max (\ln A,\ln B) \right]\, \}
\end{eqnarray}
holds. By construction the argument of the exponential function is 
negative, such that an underflow occurs when the difference between
$\min (\ln A,\ln B)$ and $\max (\ln A,\ln B)$ becomes too big, 
whereas it becomes calculable when this difference is small enough. 

To handle alternating signs one needs in addition to
equation~(\ref{lnC}) an equation for $\ln |C| = \ln |A-B|$ where
$A>0$ and $B>0$ still holds. Assuming $\ln A \ne \ln B$,
equation~(\ref{lnC}) converts for $\ln |C| = \ln |A-B|$ into
\begin{eqnarray} \label{lnAbsC}
\ln |C| & = & \max \left( \ln A,\ln B \right) \\ \nonumber
& + & \ln \{ 1 - \exp
\left[ \min (\ln A,\ln B) - \max (\ln A,\ln B) \right] \}
\end{eqnarray}
and, because the logarithm is a strictly monotone function, the sign of
$C=A-B$ is positive for $\ln A>\ln B$ and negative for $\ln A<\ln B$.

The computer implementation of equations (\ref{lnC}) and (\ref{lnAbsC})
is provided by the Fortran function {\tt addln.f} and the
Fortran subroutine {\tt addln2.f} of {\tt ForLib}, respectively.
The subroutines {\tt potts\_zln.f} and {\tt potts\_zln0.f} of
{\tt ForLib} rely on this to perform the jackknife re-weighting
analysis for various physical observables.

\subsection{Energy and specific heat calculations}

\begin{figure}[t]
 \begin{picture}(150,155)
    \put(0, 0){\includegraphics{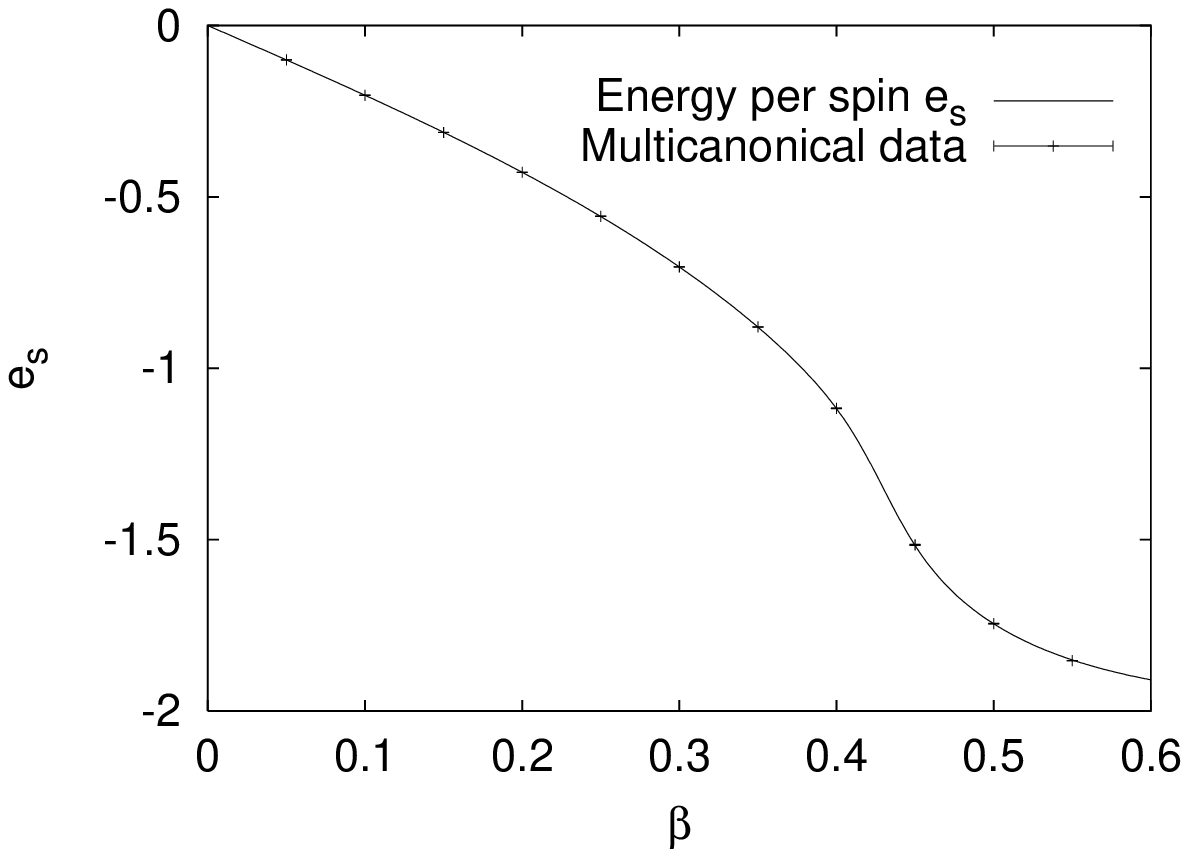}}
  \end{picture}
\caption{Energy per spin $e_s$ versus $\beta$ for the $2d$ Ising model
on an $20\times 20$ lattice, Multicanonical data are compared with the
exact result of Ferdinand and Fisher (full line), see  
the subdirectory {\tt e2\_02}.  \label{fig_2dI_es}  }
\end{figure}

\begin{figure}[t]
 \begin{picture}(150,155)
    \put(0, 0){\includegraphics{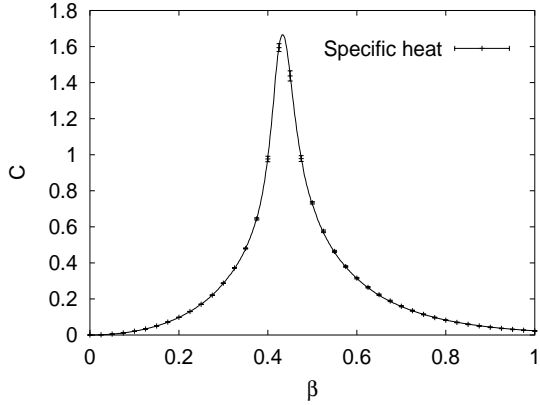}}
  \end{picture}
\caption{Specific heat versus $\beta$ for the $2d$ Ising model on a
$20\times 20$ lattice. Multicanonical data are compared with the
exact result of Ferdinand and Fisher (full line), see  
the subdirectory {\tt e2\_03}. \label{fig_2dI_C} }
\end{figure}

\begin{figure}[t]
 \begin{picture}(150,155)
    \put(0, 0){\includegraphics{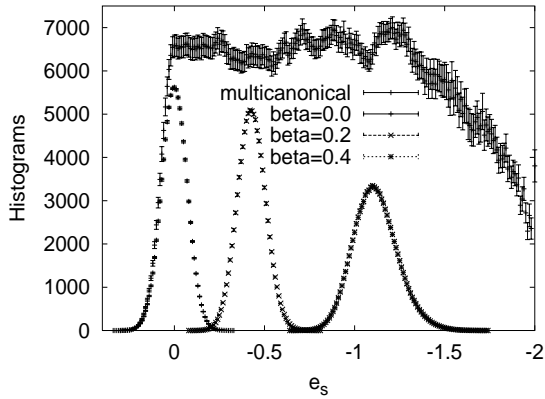}}
  \end{picture}
\caption{Energy histogram from a multicanonical simulation of the
$2d$ Ising model on a $20\times 20$ lattice together with the 
canonically re-weighted histograms at $\beta=0$, $\beta=0.2$ and 
$\beta=0.4$, see the subdirectory {\tt e2\_03}. \label{fig_2dI_muh} }
\end{figure}

\begin{figure}[t]
 \begin{picture}(150,155)
    \put(0, 0){\includegraphics{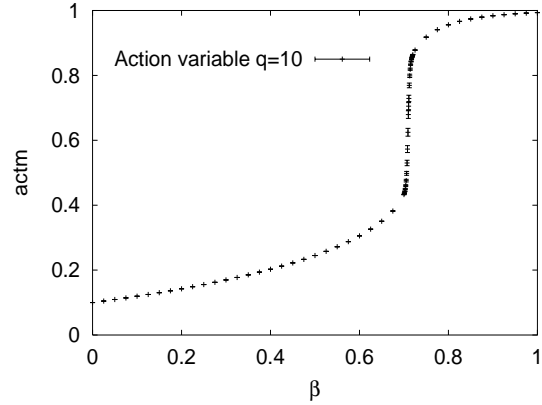}}
  \end{picture}
\caption{Multicanonical mean action variable~(\ref{actm}) data
for the $2d$ 10-state Potts model on a $20\times 20$ lattice,
see the subdirectory {\tt e2\_02}. \label{fig_2d10q_act} }
\end{figure}

\begin{figure}[t]
 \begin{picture}(150,155)
    \put(0, 0){\includegraphics{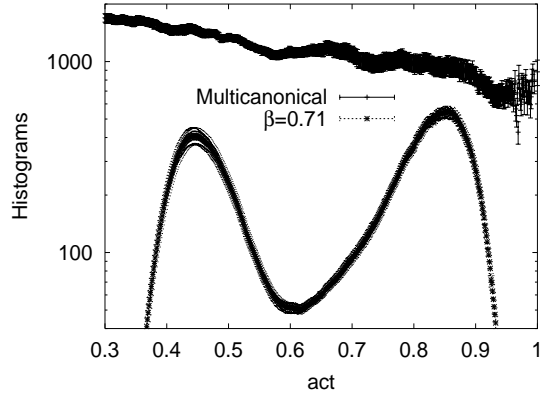}}
  \end{picture}
\caption{Action histogram from a multicanonical simulation of the
$2d$ 10-state Potts model on a $20\times 20$ lattice together
with the canonically re-weighted histograms at $\beta=0.71$, see
the subdirectory {\tt e2\_06}.  \label{fig_2d10q_muh} }
\end{figure}
 
We are now ready to analyze the MUCA data for the energy per spin of 
the $2d$ Ising model on a $20\times 20$ lattice, which we compare in
figure~\ref{fig_2dI_es} with the exact results of Ferdinand and 
Fisher~\cite{FeFi69}. The code is prepared in the 
subdirectory {\tt e2\_03}.

The same numerical technique allows us to to calculate the
{\it specific heat}, which is defined by
\begin{equation} \label{specific_heat}
 C\ =\ {d\,\hat{E}\over d\,T}\ =\ \beta^2\, \left( \,
\langle E^2 \rangle - \langle E \rangle^2 \right)\ .
\end{equation}
Figure~\ref{fig_2dI_C} compares the thus obtained MUCA data with 
the exact results of Ferdinand and Fischer~\cite{FeFi69}. The code 
is also prepared in subdirectory {\tt e2\_03}.

Figure~\ref{fig_2dI_muh} shows the energy histogram of the MUCA 
simulation together with its canonically re-weighted descendants at 
$\beta = 0$, $\beta = 0.2$ and $\beta =0.4$. The Fortran code is
prepared in the subdirectory {\tt e2\_04}. The  normalization of 
the MUCA histogram is adjusted such that it fits reasonably well
into one figure with the three re-weighted histograms.
In figure~\ref{fig_2dI_muh} it is remarkable that the error bars 
of the canonically re-weighted histograms are not just the scale 
factors times the error bars of the MUCA histogram, but in fact 
much smaller. This can be traced to be an effect of the normalization. 
The sum of each canonical jackknife histogram is 
normalized to the same number and this reduces the spread.

Relying on the $2d$ 10-state Potts model data of the run prepared in 
subdirectory {\tt e2\_02}, we reproduce in subdirectory {\tt e2\_05}
the action variable {\tt actm} results plotted in 
figure~\ref{fig_2d10q_act}.  Around $\beta = 0.71$ we observe a sharp 
increase of {\tt actm} from 0.433 at $\beta = 0.70$ to 0.864 at 
$\beta = 0.72$, which signals the first order phase transition of the 
model.

In figure~\ref{fig_2d10q_muh} we plot the canonically re-weighted 
histogram at $\beta=0.71$ together with the MUCA histogram using 
suitable normalizations, as prepared in subdirectory {\tt e2\_06}.
The ordinate of figure~\ref{fig_2d10q_muh} is on a logarithmic scale
and the canonically re-weighted histogram exhibits the double peak 
structure which is characteristic for first order phase transitions. 
The MUCA method allows then to estimate the interface tension of the 
transition by calculating the minimum to maximum ratio on larger 
lattices, see~\cite{BeNe92,Be01}.

\subsection{Free energy and entropy calculations}

At $\beta=0$ the Potts partition function $Z$ is given by equation
(\ref{K}). MUCA simulations allow for proper normalization of the 
partition function by including $\beta=0$ in the temperature
range (\ref{beta_range}). The normalized partition function
yields important quantities of the canonical ensemble, the
{\it Helmholtz free energy}
\begin{equation} \label{free_energy} 
 F\ =\ - \beta^{-1}\, \ln ( Z )
\end{equation}
and the {\it entropy}
\begin{equation} \label{entropy} 
 S\ =\ \beta\, (F - E)
\end{equation}
where $E$ is the internal energy~(\ref{iact}).

\begin{figure}[t]
 \begin{picture}(150,155)
    \put(0, 0){\includegraphics{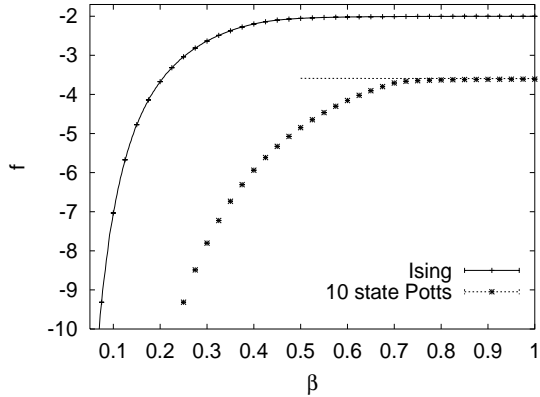}}
  \end{picture}
\caption{Free energies from multicanonical simulations of the $2d$ 
Ising and $2d$ 10-state Potts models on a $20\times 20$ lattice, 
see subdirectories {\tt e2\_03} and {\tt e2\_05}. The lines
are the exact results of Ferdinand and Fischer for the 
Ising model and the asymptotic equation~(\ref{f_as}) for the 
10-state Potts model.  \label{fig_free_energy} }
\end{figure}

\begin{figure}[t]
 \begin{picture}(150,155)
    \put(0, 0){\includegraphics{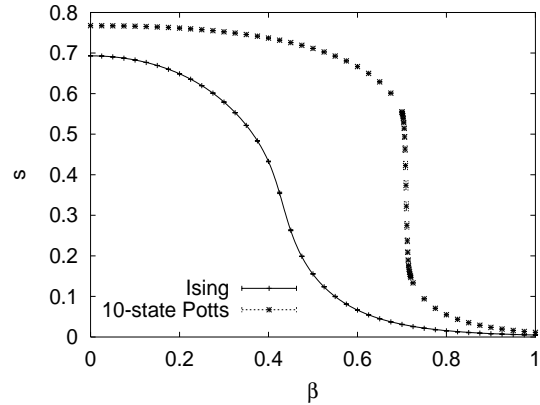}}
  \end{picture}
\caption{Entropies from multicanonical simulations of the $2d$ Ising
and $2d$ 10-state Potts models on a $20\times 20$ lattice, see the
subdirectories {\tt e2\_03} and {\tt e2\_05}. The full line is 
the exact result of Ferdinand and Fischer for the Ising model.
\label{fig_entropy} }
\end{figure}

Figure~\ref{fig_free_energy} shows the free energy density per site
\begin{equation} \label{F_density}
 f\ =\ F/N
\end{equation}
for the $2d$ Ising model as well as for the $2d$ 10-state Potts model.
The Ising model analysis is prepared in the subdirectory {\tt e2\_03} 
and the $q=10$ analysis in {\tt e2\_05}.
As in previous figures, the Ising model data are presented together
with the exact results, whereas we compare the 10-state Potts model
data with the $\beta\to\infty$ asymptotic behavior. For
large $\beta$ the partition function of our $q$-state Potts models 
approaches $q\,\exp(-2\,\beta\,d\,N+2\,\beta\,d\,N/q)$ and, therefore,
\begin{equation} \label{f_as} 
f_{\rm as} = {2\,d\over q} - 2\,d - \beta^{-1}\, {\ln (q) \over N}\ .
\end{equation}
Finally, in figure~\ref{fig_entropy} we plot for the entropy density
\begin{equation} \label{S_density}
 s\ =\ S/N
\end{equation}
the $2d$ Ising model the MUCA results together with the exact curve.  
Further, entropy data for the $2d$ 10-state Potts model are included 
in this figure.  In that case we use $s/3$ instead of $s$, such that 
both graphs fit into the same figure. The analysis code for the 
entropy is contained in the same subdirectories as used for the 
free energy.

In all the figures of this section excellent agreement between the 
numerical and the analytical results is found.

\section{Conclusions} \label{conclusions}

There are many ways to extend the multicanonical simulations
of this paper. The interested reader is simply referred to the
literature~\cite{Be01}. The purpose of this article is to serve 
a start-up kit for the computer implementation of some of 
the relevant steps. There appears to be need for this, because
to get the first program up and running appears to be a major
stumbling block in the way of using the method.

In the opinion of the author, multicanonical simulations have the 
potential to replace canonical simulations as the method of first 
choice for studies of small to medium-sized systems. As seen here,
once the recursion necessary for the first part of a MUCA simulation
is programmed, the entire thermodynamics of the system follows
from the second part of the simulation. However, the slowing down for 
larger system sizes is rather severe. 

Quite a number of similar methods exists, see~\cite{Be01} for a summary. 
A sound comparison would require that the goals of the simulations and 
their benchmarks are defined first. So far the community has not set 
such standards.

\smallskip
\noindent {\bf Acknowledgments:}
I would like to thank Alexander Velytsky for useful discussions and for
contributing figure~\ref{fig_fort}. This work was in part supported by 
the U.S. Department of Energy under the contract DE-FG02-97ER41022.

\appendix
\section{Weight Recursion} \label{Recursion}

We first discuss the weights (\ref{wspectral}). By definition, 
the microcanonical temperature is
\begin{equation} \label{T_micro}
 b (E) = {1\over T(E)} = {\partial S (E) \over \partial E}
\end{equation}
and we define the dimensionless, microcanonical free energy by
\begin{eqnarray} \label{F_micro}
 a(E) & = & {F(E)\over T(E)}\,=\,{E\over T(E)} - S(E) \\ 
      & = & b(E)\,E - S(E)\ . \nonumber
\end{eqnarray}
It is determined by relation~(\ref{T_micro}) up to an (irrelevant) 
additive constant. We consider the case of a discrete minimal 
energy $\epsilon$ and choose 
\begin{equation} \label{bE}
 b(E) = \left[ S(E+\epsilon) - S(E) \right] / \epsilon 
\end{equation}
as the definition of $b(E)$. The identity
$$ S(E) = b(E)\, E - a(E) $$
implies
$$ S(E) - S(E-\epsilon) = $$
$$ b(E) E - b(E-\epsilon) (E-\epsilon ) - a(E) + a(E-\epsilon)\ .$$
Inserting $\epsilon\, b(E-\epsilon) = S(E) - S(E-\epsilon)$ yields 
\begin{equation}
 a(E-\epsilon) =  a(E) + \left[ b(E-\epsilon)-b(E) \right]\, E
\end{equation}
and $a(E)$ is fixed by defining $a(E_{\max})=0$. Once $b(E)$ is given, 
$a(E)$ follows.

A convenient starting condition for the initial $(n=0)$ simulation is
\begin{equation} \label{b0_init}
 b^0 (E) = 0 ~~{\rm and}~~ a^0 (E) = 0\, ,
\end{equation}
because the system moves freely in the disordered phase. Other $b^0 (E)$ 
choices are of course possible. Our Fortran implementation allows for 
$b^0(E) =\beta$ with $\beta$ defined in the parameter file {\tt mc.par}. 

The energy histogram of the $n^{th}$ simulation is given by $H^n(E)$.
To avoid $H^n(E)=0$ we replace for the moment
\begin{equation} \label{hatH}
H^n(E)\to {\hat H^n(E)} = \max\, [h_0,H^n(E)]\, ,
\end{equation}
where $h_0$ is a number $0 < h_0 < 1$.  Our final equations allow for 
the limit $h_0 \to 0$. In the following subscripts $_0$ are used to 
indicate quantities which are are not yet our final estimators from the
$n^{th}$ simulation. We define
$$w^{n+1}_0(E)=e^{-S^{n+1}_0(E)}=c\, {w^n(E) \over \hat{H}^n(E)}\, ,$$
where the (otherwise irrelevant) constant $c$ is introduced to ensure 
that $S^{n+1}_0(E)$ is an estimator of the microcanonical entropy
\begin{equation} 
 S^{n+1}_0(E) = - \ln c+S^n(E)+\ln \hat{H}^n(E)\ .
\end{equation}
Inserting this relation into (\ref{bE}) gives
\begin{equation} \label{b0} 
 b^{n+1}_0 (E) = 
\end{equation}
$$ b^n (E) + [ \ln \hat{H}^n(E+\epsilon) - 
  \ln \hat{H}^n(E) ] / \epsilon\, .$$
The estimator of the variance of $b^{n+1}_0(E)$ is obtained from 
$$\sigma^2 [ b^{n+1}_0(E)] = \sigma^2[ b^n (E) ] + $$
$$  \sigma^2 [ \ln \hat{H}^n(E+\epsilon)] / \epsilon + 
  \sigma^2[  \ln \hat{H}^n(E) ] / \epsilon\ . $$
Now $\sigma^2[b^n (E)]=0$ as $b^n(E)$ is the fixed function used in
the $n^{th}$ simulation and the fluctuations are governed by
the sampled histogram $H^n=H^n(E)$
$$ \sigma^2[ \ln (\hat{H}^n )] = $$ 
$$ \sigma^2[ \ln (H^n)] = 
\left[ \ln (H^n + \triangle H^n) - \ln (H^n) \right]^2 $$
where $\triangle H^n$ is the fluctuation of the histogram, which
is known to grow with the square root of the number of entries
$\triangle H^n \sim \sqrt{H^n}$. Hence, under the assumption that
autocorrelation times of neighboring histogram entries are identical, 
the equation
\begin{equation} \label{sigma2_b} 
 \sigma^2[ b^{n+1}_0(E)]
={c'\over H^n(E+\epsilon)}+{c'\over H^n(E)}
\end{equation}
holds, where $c'$ is an unknown constant. The assumption would
be less strong if it were made for the energy-dependent
acceptance rate histogram instead of the energy histogram. In
the present models the energy dependence of the acceptance rate
is rather smooth between nearest neighbors and there is less 
programming effort when using only energy histograms. 
Equation (\ref{sigma2_b}) shows that 
the variance is infinite when there is zero statistics for either
histogram, $H^n(E)=0$ or $H^n(E+\epsilon)=0$. The statistical 
weight for $b^{n+1}_0 (E)$ is inversely proportional to its variance and 
the over-all constant is irrelevant. We define
\begin{eqnarray} \label{gn0}
g^n_0 (E) & = & {c'\over \sigma^2[ b^{n+1}_0(E)]} \\ \nonumber
& = & {H^n (E+\epsilon)\ H^n (E) \over H^n (E+\epsilon) + H^n (E)} 
\end{eqnarray}
which is zero for $H^n(E+\epsilon)=0$ or $H^n(E)=0$.
The $n^{th}$ simulation is carried out using $b^n (E)$. It is
now straightforward to combine $b_0^{n+1} (E)$ and 
$b^n (E)$ according to their respective statistical weights
into the desired estimator:
\begin{equation} \label{bn}
b^{n+1} (E) = \hat{g}^n (E)\,  b^n (E) +
                \hat{g}^n_0 (E)\, b^{n+1}_0 (E)\, ,
\end{equation}
where the normalized weights
\begin{equation} \label{hatgn0}
 \hat{g}^n_0 (E) = {g^n_0(E) \over g^n(E) + g^n_0 (E)}
\end{equation}
and
\begin{equation} \label{hatgn}
 \hat{g}^n (E) = 1 - \hat{g}^n_0 (E) 
\end{equation}
are determined by the recursion 
\begin{equation} \label{gn_recursion}
g^{n+1} (E) = g^n(E) + g^n_0(E),\ g^0(E)=0\, .
\end{equation}
We can eliminate $b^{n+1}_0 (E)$ from equation~(\ref{bn}) by
inserting its definition (\ref{b0}) and get
\begin{equation} \label{recursion}
 b^{n+1}(E) = b^n (E) +
\end{equation}
$$ \hat{g}^n_0(E) \times
 [ \ln \hat{H}^n(E+\epsilon)-\ln \hat{H}^n(E)] / \epsilon\ . $$
Notice that it is now save to perform the limit $h_0\to 0$.
Finally, equation~(\ref{recursion}) can be converted into a direct 
recursion for ratios of the weight factor neighbors. We define
\begin{equation} \label{wrat}
R^n (E) = e^{\epsilon b^n (E)} = {w^n(E) \over w^n (E+\epsilon)}
\end{equation}
and get
\begin{equation} \label{wrat_recursion}
R^{n+1} (E) = R^n (E)\, \left[ {\hat{H}^n (E+\epsilon) \over
              \hat{H}^n (E) } \right]^{\hat{g}^n_0(E)}\, .
\end{equation}

\end{document}